\documentclass[aps,twocolumn]{revtex4}

\usepackage{bm}
\usepackage{amsmath}
\usepackage{amssymb}    
\usepackage{amsbsy}     
\usepackage{amstext}   
\usepackage{array}
\usepackage{epsfig}

\begin{document}

\title{A two-channel R-matrix analysis of magnetic field induced Feshbach resonances}

\author{Nicolai Nygaard}
\email{nygaard@phys.au.dk}
\affiliation{Danish National Research Foundation Center for Quantum Optics, Department of Physics and Astronomy, University of Aarhus,  
DK-8000 {\AA}rhus C, Denmark}        
\author{Barry I. Schneider}
\email{bschneid@nsf.gov}
\affiliation{Physics Division, National Science Foundation, Arlington,
Virginia 22230 and Electron and Optical Physics Division, National Institute
of Standards and Technology, Gaithersburg, MD 20899} 
\author{Paul S.~Julienne}
\email{paul.julienne@nist.gov}
\affiliation{Atomic Physics Division, National Institute
of Standards and Technology, Gaithersburg, MD 20899}       

\date{\today}

\begin{abstract}

A Feshbach resonance arises in cold atom scattering due to the complex interplay between several coupled channels. However, the essential physics of the resonance may be encapsulated in a simplified model consisting of just two coupled channels. In this paper we describe in detail how such an effective Feshbach model can be constructed from knowledge of a few key parameters, characterizing the atomic Born-Oppenheimer potentials and the low energy scattering near the resonance. These parameters may be obtained either from experiment or full coupled channels calculations. Using R-matrix theory we analyze the bound state spectrum and the scattering properties of the two-channel model, and find it to be in good agreement with exact calculations.

\end{abstract}

\maketitle

\section{Introduction}
\label{Introduction}

A Feshbach resonance can occur in collision events, when a closed channel bound state of a specific electronic level of an atom is coupled to the scattering continuum of an open channel of another electronic level of the atom ~\cite{Feshbach1958_1962}. If the energy of the discrete closed channel state  is very close to that of the scattering particles, the cross section is resonantly enhanced, due to the temporary capture of the particles in the quasi-bound state. 
Feshbach resonances would be of limited importance in cold atom scattering if not for the ability to control the energy of the bare resonance state, and hence the resonance position, through the application of a static magnetic field~\cite{Tiesinga1993,Moerdijk1995}. In this context the open and closed channels, in the separated atom limit, correspond  to pairs of alkali atoms in different combinations of eigenstates of the hyperfine and Zeeman Hamiltonian with total angular momentum projection quantum numbers $m_{F,1}$ and $m_{F,2}$. The entrance channel is in an $s$-wave orbital state (relative angular momentum quantum number $l=0$), while the closed channels may be in any partial wave, with $m_l$ indicating the projection of the relative orbital angular momentum. All channels have the same projection of the total angular momentum along the magnetic field axis, $m_F=m_{F,1}+m_{F,2}+m_l$, and since their magnetic moments are different, they experience a relative Zeeman shift in an applied magnetic field. Hence a closed channel bound state may be shifted in energy relative to the open channel threshold, bringing it into resonance. The coupling between the entrance channel and $s$-wave closed channels arises from the central part of the molecular potential, which only depends on the magnitude of the total electron spin. Therefore, the interatomic  potential has short-range off-diagonal terms proportional to the difference between the triplet and singlet potentials. Closed channels of higher angular momentum ($l>0$) can be coupled to the entrance channel through the spin-dipole inteaction~\cite{Stoof1988}.

The Feshbach coupling gives rise to a dispersive variation with magnetic field of the scattering length, well described by~\cite{Moerdijk1995}
\begin{equation}
\label{a_vs_B}
a(B) = a_{\rm{bg}} \left[ 1 - \frac{\Delta B}{B-B_0} \right].
\end{equation}
This is associated with a bound molecular state of the coupled channels problem crossing threshold at $B=B_0$. The width of the resonance is characterized by $\Delta B$, and $a_{\rm{bg}}$ is the limiting value of the scattering length for magnetic fields far removed from $B_0$. In ultracold atomic gases the $s$-wave scattering length gives a complete characterization of the interactions, and (\ref{a_vs_B}) therefore implies an unmitigated control of the sign and strength of the effective interparticle potential. This has facilitated the experimental investigation of the crossover between loosely bound Cooper pairs and a Bose-Einstein condensate of true dimers in a trapped Fermi gas~\cite{Ohara2002,Regal2004,Zwierlein2004,Bartenstein2004,Kinast2004,Bourdel2004}.
 
The coupled channels description of atomic scattering may be solved in full generality once the adiabatic Born-Oppenheimer (BO) curves for the singlet and triplet potentials are known, and gives accurate predictions for the location of the experimentally observed Feshbach resonances. However, there is decided usefulness in the availability of simplified model descriptions, which can be included effortlessly in many-body~\cite{Goral2004,Koehler2004,Simonucci2005,Szymanska2005} and three-body~\cite{Stoll2005} treatments, or are more amenable to analysis of time-dependent experiments~\cite{Mies2000,Koehler2003,Goral2004}.  Fortunately, in most atoms Feshbach resonances occur at well separated magnetic fields. Hence, a natural starting point for a simplified description is to consider an isolated resonance. 

In this paper, we demonstrate that it is possible to reduce the full
multi-channel, low-energy scattering problem of two alkali atoms to an effective 
two-channel problem for a single Feshbach resonance~\cite{Mies2000,Goral2004}. The model consists of a single closed channel containing a bound state, the bare resonance state, which interacts with the scattering continuum in the open channel. In this sense it is equivalent to Fano's theory of resonances arising from a discrete state embedded in the continuum~\cite{Fano1961}, but the bare resonance state in the two-channel model is a construct, which may represent the interactions between several closed channels in the full coupled channels problem.
This model is capable of capturing the relevant parts of the full molecular physics of the resonance, provided it is constructed to comply with a small set of low energy parameters. Section~\ref{2channel_model} gives an explicit prescription for constructing the two-channel pseudopotentials. 

The R-matrix formalism, which is presented in Section~\ref{R-matrix_solution} is an elegant and efficient method of calculating both bound state and scattering properties. It expresses the logarithmic derivative of the wavefunction at some cutoff radius, beyond which the asymptotic solutions of the Schr{\"o}dinger equation are known, in terms of the solutions of an eigenvalue problem. In Section~\ref{Numerics} we discuss briefly the specific numerical implementation of the method we have used in this paper. We argue that a Discrete Variable Representation of the Hamiltonian combined with a grid based on a Finite Element partinioning of space defines a numerical procedure, which is efficient in the evaluation of matrix elements and capable of handling the two widely separated length scales associated with the molecular and low energy scattering physics. 

Section~\ref{Results} gives the results of our calculations for a particular Feshbach resonance. In addition, we give the parameters defining our two-channel model for several other resonances of interest. In all cases we have ensured that the two-channel model gives an accurate representation of the molecular physics, by comparing with full coupled channels calculations. We conclude our findings in Section~\ref{Conclusions}, where we also contrast our pseudopotential approach with that of other authors.
 
%In essence, we are presented with a scattering problem in which a single open channel level in one BO electronic state, interacts with a closed channel level in another BO state to produce a resonance when the (short-range) coupling between the two BO states is introduced.  The presence of an external magnetic field provides a mechanism to "tune" the interaction from one that supports loosely bound Cooper pairs to one that supports a molecular Bose-Einstein condensate. 

\section{The two-channel model of Feshbach resonances}
\label{2channel_model}

In the case of an isolated Feshbach resonance it is possible to reduce the full coupled channels problem to an effective two-channel model with one open channel (1) and one closed channel (2) with a bound state, which interacts with the open channel scattering continuum to produce the resonance. Such a model is capable of capturing the essential physics of the full multi-channel scattering calculation over a wide range of energies. In this section we set up the two-channel model, following the procedure outlined in Refs.~\cite{Mies2000,Koehler2003}, and discuss how a correspondence may be established between its parametrization and low energy scattering observables, available either through experiments or more exact calculations.  

The coupled two-channel Hamiltonian describing the relative motion of the particles is chosen to be 
\begin{equation}
\label{H2B}
{\bf H}_{2B}=\left(  
\begin{array}{cc}
-\frac{\hbar^2}{2\mu} \frac{d^2}{dr^2} + V_1(r) & W(r) \\
W(r) & -\frac{\hbar^2}{2\mu} \frac{d^2}{dr^2} + V_2(r;B)
\end{array}
\right),
\end{equation}
where $\mu$ is the reduced mass.  We take the zero of energy to correspond to the energy of the dissociated atoms in channel 1 at all magnetic field values. Correspondingly, the only magnetic field dependent component of the problem is the energy offset of the second channel. 
In the absence of any coupling between the channels the eigenstates of the two-channel problem is given by the diabatic states $(\phi_{1,n},0)$, and $(0,\phi_{2,n})$, where $\phi_{i,n}$ are the eigenstates of the Hamiltonian for channel $i$. The coupling mixes these states, and the resulting dressed (or adiabatic) states will in general have components in both channels.
Our intention is to solve the coupled channels eigenvalue problem 
\begin{equation}
\label{eigenvalue_problem}
({\bf H}_{2B}-E{\bf I})
\left(
\begin{array}{cc}
\psi_1(r) \\
\psi_2(r)
\end{array}
\right)=0,
\end{equation}
for the components of the wavefunction in each of the two channels, both for the discrete part of the spectrum and for the scattering states. Here ${\bf I}$ is the two by two identity matrix. 

\begin{figure}[htbp]
\begin{center}
\includegraphics[scale=0.4]{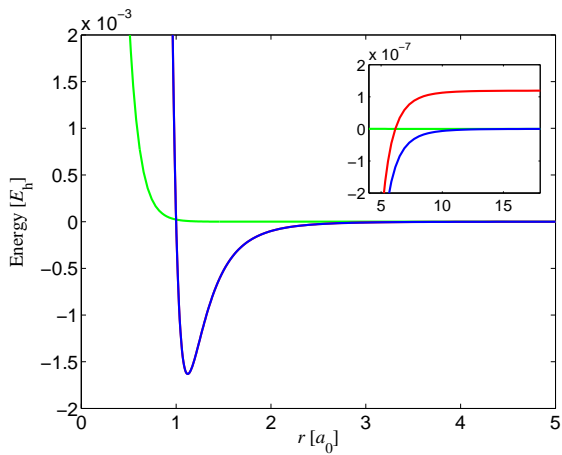}
\caption{Diagonal and off-diagonal potential curves for the two-channel model representation of
the $^{40}$K Feshbach resonance at $B_0=202.15$ G (1 G=0.0001 T). The open potential, $V_1(r)$ is plotted in blue, while the closed channel potential at zero field, $V_2(r,0)$, is represented by the red curve. The diagonal potentials are indistinguishable on this scale. The interchannel coupling, $W(r)$, is plotted in green. The inset shows the long-range behavior of the potentials, and the threshold energy $E_{\rm{th}}$ is evident as the offset between the asymptotic values of the open and closed channel potentials. 1 $E_h=4.359744\times 10^{-18}$ J.}
\label{potentials}
\end{center}
\end{figure}

The BO potentials are dominated by the van der Waals interactions at large internuclear separation, vanishing as $r^{-6}$, as $r\rightarrow \infty$. At short distances there is a steep potential barrier. Any potential we choose with these characteristics will give an accurate representation of the molecular physics at low energies. For simplicity we choose a Lennard-Jones potential:
\begin{equation}
\label{LJ}
V_1(r) = -4\epsilon \left[  \left(\frac{\sigma}{r} \right)^6  - \left(\frac{\sigma}{r}\right)^{12} \right].
\end{equation}
As in Refs.~\cite{Mies2000,Koehler2003} we take the potential in both channels to have the same form, but with the closed channel potential shifted upwards in energy, such that its threshold is at $E_{{\rm{th}}}+\Delta\mu B$, where $\Delta\mu$ denotes the difference in magnetic moment between the separated atoms and the bare resonance state. 
This simplifies the numerical analysis, and does not seem to have a noticeable effect on the results.
Hence, we have
\begin{equation}
\label{closed_channel}
V_2(r;B) = V_1(r) + E_{{\rm{th}}} +\Delta\mu B.
\end{equation}
Finally, we represent the coupling between the two channels by an off-diagonal potential $W(r)$. Again, the specific form of this potential is arbitrary, but it should be a short-range function, since the inter-channel coupling is only significant at small distances. Specifically, we use an exponential form
\begin{equation}
\label{coupling}
W(r) =W_0 e^{-r/a_0},
\end{equation} 
where $a_0=0.529177\times 10^{-10} {\rm{m}}$ is the Bohr radius. We thus end up with a five parameter model for the Feshbach resonance: $\epsilon$ and $\sigma$ characterize the depth and shape of the diagonal potentials, respectively, while the off-diagonal coupling is of strength $W_0$. Finally, the offset between the threshold energies of the two channels for a given magnetic field strength is determined by  $\Delta\mu$ and $E_{{\rm{th}}}$. These five parameters are determined by matching the low energy scattering properties of the two-channel model with those of full coupled channels calculations or experimental observables. The model potentials are illustrated in Fig.~\ref{potentials} for the case of $^{40}$K. 

To make this connection explicit we make use of analytical results for low energy scattering in inverse power law potentials derived by Flambaum {\textit{et al}}.~\cite{Flambaum1999}. For the open channel, which determines the far from resonance, or background, scattering properties, we require that our model potential $V_1(r)$ supports $N_s$bound states and that its scattering length equals the background value, $a_{\rm{bg}}$, which appears in (\ref{a_vs_B}). We only require that $N_s$ is of the order of the number of bound states in the actual potential, since when $N_s \gg 1$, the near-threshold properties are essentially independent of its actual magnitude. Given the atomic mass and the $C_6$ coefficient, which is well known from the asymptotic scattering properties, we may unambiguously construct a model potential meeting these constraints.  Both $N_s$ and $a_{\rm{bg}}$ are related to the WKB phase $\Phi$ evaluated at zero energy:
\begin{eqnarray}
a_{\rm{bg}} &=& \bar{a} \left[  1-\tan(\Phi -\pi/8) \right], \\
N_s &=& \left[  \frac{\Phi}{\pi}-\frac{5}{8} \right] +1,
\end{eqnarray}
where $[ \ ]$ is the integer part, and the mean scattering length (or characteristic scale length of the van der Waals potential) $\bar{a}=\sqrt{4\mu C_6}\Gamma(\frac{3}{4})/\Gamma(\frac{1}{4})\hbar$ is governed solely by the asymptotic behavior of the potential. The second equation is necessary to completely determine the semiclassical phase, since the first equation can only be solved for $\Phi$ modulo $\pi$. Following Flambaum {\textit{et al}}. we can then find the relationship between the $C_{12}$ and $C_6$ coefficients
\begin{equation}
C_{12}  = \left( \frac{\sqrt{2\pi\mu} \Gamma(\frac{1}{3})C_6^{5/6}}{10\Gamma(\frac{5}{6})\Phi\hbar} \right)^3.
\end{equation} 
Knowing both the $C_6$ and $C_{12}$ coefficients of our model potential, the range and depth parameters are uniquely determined:
\begin{eqnarray}
\sigma &=& \left( \frac{C_{12}}{C_6} \right)^{1/6}, \\
\epsilon &=& \frac{C_6}{4\sigma^6}.
\end{eqnarray}

The bare resonance state is a particular unit normalized bound state of closed channel potential:
\begin{equation}
\left[ -\frac{\hbar^2}{2\mu} \frac{d^2}{dr^2} +V_2(r;0) \right] \phi_{\rm{res}}(r) = E_{\rm{res}}(0) \phi_{\rm{res}}(r).
\end{equation}
We arbitrarily choose $\phi_{\rm{res}}(r)$ to be the second to last bound state of the closed channel potential, such that $E_{\rm{res}}(0)=E^{(2)}_{-2}+E_{\rm{th}}$. Here $E^{(2)}_{-2}$ indicates the binding energy of the bare resonance state relative to the threshold in the {\textit{closed}} channel. This is calculated by diagonalizing $-(\hbar^2/2\mu)(d^2/dr^2)+V_1(r)$. 
A calculation of the scattering properties of the two-channel model~\cite{Goral2004,Julienne2004} reveals that the magnetic field dependent scattering length (\ref{a_vs_B})
can be related to the strength of the model coupling potential via
\begin{equation}
\label{DeltaB}
\Delta B = \lim_{k\rightarrow 0}  \frac{\Gamma(E)}{2ka_{\rm{bg}}\Delta\mu},
\end{equation}
where $E=\hbar^2k^2/2\mu$.
For $E>0$ the decay width of the quasi-bound molecule giving rise to the resonance is given by ~\cite{Fano1961,Feshbach1958_1962}
\begin{equation}
\label{Gamma}
\Gamma(E) = 2\pi |\langle  \phi_{\rm{res}} | W | \phi_{E} \rangle|^2.
\end{equation}
Here $\phi_E(r)$ is the energy normalized outgoing wave scattering state at energy $E$ in the presence of the open channel potential alone
\begin{equation}
\left[ -\frac{\hbar^2}{2\mu} \frac{d^2}{dr^2} +V_1(r) \right] \phi_{E}(r)=E \phi_{E}(r).
\end{equation}
Its asymptotic form determines the background contribution to the scattering phase shift, $\delta_{\rm{bg}}(E)$
\begin{equation}
\phi_E(r) \rightarrow \sqrt{\frac{2\mu}{\pi\hbar^2 k}} \sin\left(kr+\delta_{\rm{bg}}(E)\right).
\end{equation}
The background scattering length follows from $a_{\rm{bg}}=-\tan \delta_{\rm{bg}}/k$ in the limit of $E\rightarrow 0$. 
We specify below how the matrix element of the coupling operator in (\ref{DeltaB}) is calculated in the R-matrix formalism. Since the scattering length $a(B)$ is zero at the magnetic field $B=B_0+\Delta B$, the width of the resonance can be either measured or obtained from multi-channel calculations. Hence the strength of the interchannel coupling function may calculated using (\ref{DeltaB}), assuming the difference in magnetic moment of the two channels $\Delta\mu$ is known. We will explain below how this quantity may be obtained. 

The measured or calculated position of the Feshbach resonance, $B_0$, is shifted away from the field value $B_{\rm{res}}$, where the bare resonance state crosses zero energy, due to the coupling between the channels . An accurate analytic estimate of this shift is given by multi-channel quantum defect theory~\cite{Julienne2004}
\begin{equation}
B_0-B_{\rm{res}} = \Delta B \frac{x(1-x)}{1+(1-x)^2}, \ \ \ x = a_{\rm{bg}}/\bar{a} .
\end{equation}
This provides a link to the threshold energy, $E_{\rm{th}}$. Since the variation of the bare resonance state with magnetic field is given by $E_{\rm{res}}(B)=E_{\rm{res}}(0)+\Delta\mu B$, it follows from $E_{\rm{res}}(B_{\rm{res}})=0$ that
\begin{equation}
E_{\rm{th}} = -(\Delta\mu B_{\rm{res}} + E_{-2}). 
\end{equation}

Finally, we address the derivative of the bare resonance state energy with respect to the magnetic field. From the Fano theory of resonant scattering~\cite{Fano1961} we know that a bound state embedded in the continuum gives rise to a resonant contribution to the phase shift
\begin{equation}
\delta_{\rm{res}}(E,B) =  \tan^{-1} \frac{\Gamma(E)/2}{E-E_{\rm{res}}(B)+\Delta(E)}.
\end{equation}
The shift of the resonance position away from the energy of the bare resonance state is~\cite{Fano1961,Feshbach1958_1962} 
\begin{equation}
\Delta(E) = {\mathcal{P}} \int \frac{dE'}{2\pi} \frac{\Gamma(E')}{E-E'}.
\end{equation}
Importantly, this shift is independent of the magnetic field. Hence, we expect that for magnetic fields sufficiently far from $B_0$ that the resonance energy is unaffected by threshold effects, the position of the scattering resonance may be fitted with a linear function of the magnetic field, with a slope of $\Delta\mu$. This defines a procedure for determining the last parameter of our two-channel model.

As we shall demonstrate below, this procedure for constructing an effective two-channel model yields predictions, which accurately reproduce the results of a more rigorous calculation for both the continuum {\textit{and}} the bound state spectra.

\section{Solution using the R-matrix method}
\label{R-matrix_solution}

The R-matrix method~\cite{P.Burke_C.Noble_V.Burke,B.Schneider_R-Matrix} has been demonstrated over the past thirty-five years to be a general and quite powerful ab-initio approach to a wide class of atomic and molecular collision problems.  The essential idea is to divide space into two or possibly more physical regions.  In each of these regions the time independent Schr\"odinger equation may be solved using techniques designed to be optimal to describe the important physical properties of that region.  The solutions and their derivatives are then matched at the region interfaces.  A simple example to illustrate this idea is electron scattering from atomic targets.  In this case, when the electron is close to the other atomic electrons, there are strong electrostatic, exchange and correlation effects.  Outside this internal region, the electron being scattered is subject to purely long-range electrostatic forces.  It then becomes possible to expand the wavefunction in the internal region in a discrete set of eigenstates using the full power of standard bound state approaches.  The internal wavefunction is then matched to known or easily computed solutions of the long-range potentials on the surface of the sphere bounding the two regions.
\par
For the two channel problem considered in this paper, using the definitions in Section \ref{2channel_model}, we define the R-matrix eigenstates, \{$\psi_{c,n}(r)$ ($c=1,2$)\}, inside a sphere of radius of $r=a$,
\begin{equation}
\label{2_channel_R_matrix_eqns}
({\bf H}_{2B} + L{\bf I} -E_n{\bf I})
\left(
\begin{array}{c}
\psi_{1,n}(r) \\
\psi_{2,n}(r)
\end{array}
\right)
=0,
\end{equation}
where $L$, the Bloch operator~\cite{Bloch} is,
\begin{equation}
L = \frac{\hbar^2}{2\mu} \delta ( r - a ) \frac{\partial}{\partial r}.
\end{equation}
The function of the Bloch operator is to ensure that any non-zero contributions to matrix elements arising from the kinetic energy operator on the surface of the R-matrix sphere are exactly cancelled by contributions from the Bloch operator.  This ensures that the operator $ H + L$ is Hermitian and that (\ref{2_channel_R_matrix_eqns}) has only real eigenvalues.  We use a notational convention where the bound states ($E<0$) have $n<0$, while the discrete continuum states are labeled with $n=0,1,2,\ldots$, their number limited by the finite size of the basis.
The R-matrix boundary $a$ is chosen large enough that the influence of the potentials at $r=a$ is negligible, such that  
the R-matrix eigenstates match unto free particle solutions. These eigenfunctions are then used to expand the solution to the two channel problem at an arbitrary energy, $E$, as
\begin{equation}
\label{2_channel_internal_solution}
\left(
\begin{array}{c}
\psi_{1,E}(r) \\
\psi_{2,E}(r)
\end{array}
\right)
                    = \sum_n a_n
\left( 
\begin{array}{c}
\psi_{1,n}(r) \\
\psi_{2,n}(r)
\end{array} 
\right ).
\end{equation}
The coefficients, $a_n$, are determined by matching the solution and its first derivative to appropriate asymptotic forms at $r=a$.  The boundary $a$ is chosen sufficiently large that when both channels are closed, corresponding to true vibrational bound states of the coupled system with $E<0$, the wavefunction has decayed to zero at the boundary.  This ensures that the R-matrix eigenstates reproduce the bound state vibrational spectrum of the molecules precisely.  For the scattering states the matching conditions give,
\begin{equation}
\label{matching}
\left(
\begin{array}{c}
\psi_{1,E}(a) \\
\psi_{2,E}(a)
\end{array}
\right)  
               = 
\left(
\begin{array}{cc}
{\mathcal{R}}_{11} & {\mathcal{R}}_{12} \\
{\mathcal{R}}_{12} & {\mathcal{R}}_{22}
\end{array}
\right )
\left (
\begin{array}{c}
\psi'_{1,E}(a) \\
\psi'_{2,E}(a) 
\end{array}
\right),
\end{equation}
where
\begin{equation}
\label{R-matrix_element}
{\mathcal{R}}_{c,c'} = \frac{\hbar^2}{2\mu} \sum_n \frac{ \psi_{c,n}(a) \psi_{c^{\prime},n}(a) } { E_n - E },  \quad \quad c=1,2,
\end{equation}
is the R-matrix for our problem, and $\psi'_{c,E}(a)$ is the derivative of the channel wavefunction at the boundary. It is convenient to define the matrix
\begin{equation}
{\bf k}^{-1/2} = 
\sqrt{\frac{2\mu}{\pi\hbar^2}} 
\left(  
\begin{array}{cc}
1/\sqrt{k_1} & 0 \\
0 & 1/\sqrt{k_2}
\end{array}
\right),
\end{equation}
which determines the energy normalization of the scattering states. Here 
\begin{equation}
k_1=\sqrt{\frac{2\mu E}{\hbar^2}}, \ 
k_2=\sqrt{\frac{2\mu (E-E_{\rm{th}}-\Delta\mu B)}{\hbar^2}}.
\end{equation}

In the case where channel $1$ is open and channel $2$ is closed, the asymptotic scattering state for $s$-wave collisions can be written as
\begin{equation}
\label{2_channel_external_solution_oc}
\left(
\begin{array}{c}
\psi_{1,E}(r) \\
\psi_{2,E}(r)
\end{array}
\right)  
%\xrightarrow{r \rightarrow \infty} 
\sim
\left(
\begin{array}{c}
\sqrt{\frac{2\mu}{\pi\hbar^2 k_1}} [\sin (k_1r) + K \cos(k_1r) ] \\
A \exp( - \kappa r)/r
\end{array}
\right),
\end{equation}
where $\kappa=ik_2$, $A$ is an irrelevant normalization constant, and the phase shift is given by $\tan \delta(E)=K$. Upon use of the R-matrix matching condition, the phase shift in the open channel is found to be
\begin{equation}
\label{tan_delta}
\tan \delta(E) = \frac{k_1 \bar{\mathcal{R}}_{11} \cos(k_1 a)- \sin(k_1 a)}{\cos(k_1 a) + k_1 \bar{\mathcal{R}}_{11}\sin(k_1 a)},
\end{equation}
where we have defined an effective open channel R-matrix element, $\bar{\mathcal{R}}_{11}={\mathcal{R}}_{11} - |{\mathcal{R}}_{12}|^2\kappa/(1+{\mathcal{R}}_{22}\kappa)$. In the absence of coupling, $W_0=0$, we have $\bar{\mathcal{R}}_{11}={\mathcal{R}}_{11}$ and $\psi_{1,E}(r)=\phi_E(r)$.

When both channels are open, we must allow for the possibility that the particles may enter the collision region in one channel and exit in the other. This is described by a $2\times 2$ K-matrix, ${\bf K}$, and the asymptotic scattering state becomes in a general matrix notation
\begin{equation}
\label{2_channel_external_solution_oo}
{\bm \psi}_E(r) 
%\xrightarrow{r \rightarrow \infty} 
\sim {\bf k}^{-1/2} [{\bf f}(r) + {\bf g}(r){\bf K}]. 
\end{equation}
The off-diagonal elements of ${\bm \psi}_E(r)$ describe the cases where the incoming channel and outgoing channel differ. The matrices $f_{ij}(r)=\sin(k_i r)\delta_{ij}$ and $g_{ij}(r)=\cos(k_i r)\delta_{ij}$ give the regular and irregular solutions in the asymptotic region~\cite{Mott1965}. The R-matrix matching at $r=a$ then gives a linear system of equations for the multi-channel K-matrix:
\begin{equation}
[{\bf k}^{-1/2} {\bf g}(a) - {\bm {\mathcal{R}}} {\bf k}^{-1/2} {\bf g'}(a)] {\bf K} 
= [{\bm {\mathcal{R}}} {\bf k}^{-1/2} {\bf f'}(a) - {\bf k}^{-1/2} {\bf f}(a) ]. 
\end{equation}
The S-matrix can be found from 
\begin{equation}
{\bf S} = \frac{{\bf I} + i{\bf K}}{{\bf I}-i{\bf K}}.
\end{equation}
The multi-channel S-matrix satisfies the unitarity condition $\sum_i |S_{ij}|^2=1$ for all $j$.

A great advantage of the R-matrix method can be seen from (\ref{R-matrix_element}).  The explicit energy dependence enables the R-matrix to be calculated quite efficiently with energy.  In the presence of resonances, the phase shift varies rapidly with energy but the corresponding R-matrix is constructed easily from energy independent parts.  In addition, the R-matrix poles are often very good approximations to the positions of the resonances.  The widths may be extracted in a variety of ways, the simplest being computing the phase shift as a function of energy and fitting the result to a Breit-Wigner form plus background term.  This makes R-matrix theory ideal for the study of magnetic field induced Feshbach resonances. 

The original derivation of the R-matrix method by Wigner and Eisenbud~\cite{WigEis} used R-matrix states satisfying a fixed, zero derivative boundary condition at the R-matrix radius.  While this is perfectly acceptable formally, it is a very slowly convergent process numerically as a consequence of the fact that the scattering wavefunction at energy $E$ satisfies a different boundary condition at the R-matrix radius.  Thus, the use of fixed boundary condition states, can only be correct in the limit of a complete expansion basis.  The most appropriate way to look at the problem from a variational viewpoint is that we are finding the Rayleigh-Ritz solution to the self-adjoint operator $H + L$.  By chosing a basis which does not satisfy a fixed derivative condition at the surface, convergence of the solution is quickly achieved with an incomplete basis.  It is possible to devise an approximate correction procedure~\cite{Buttle} to account for the infinity of neglected levels using a fixed boundary condition basis.  In the approach above, the use of the Bloch operator eliminates the need for specifying a specific derivative condition on the basis.  In fact, basis functions with arbitrary derivatives may be used as an expansion set.  This leads to a very rapid convergence of the S-matrix as a function of the basis set size.  The higher lying levels have little physical significance individually, but their spectral sum is effectively converging to the R-matrix.  In addition, it can be shown that truncating the very high lying R-matrix levels has a negligible effect on the low-energy scattering properties.  This is in contrast to some other approaches~\cite{Siegert} where the elimination of a single state gives completely erroneous results.

\section{Numerical approach}
\label{Numerics}

In the R-matrix method as outlined in the last section, the critical element is the computation of the R-matrix eigenvalues and eigenvectors.  The Bloch operator formalism enables the use of powerful variational methods, which do not require the underlying basis set to satisfy any specific boundary condition.  This produces a very rapid convergence of the results with respect to basis set size.  The choice of primitive basis in which to expand the R-matrix eigenstates is, of course, arbitrary.  In practice many choices are eliminated on the basis of computational convenience or efficiency.  For the problem under consideration in this paper, there is the additional constraint that the basis accurately describe both the two-body vibrational bound states as well as the very low energy continuum states. For the calculations presented here this requires a highly accurate description of the wavefunction to internuclear distances of approximately 1000$a_0$.  The Finite Element Discrete Variable (FEDVR) basis, described in other publications ~\cite{ResMcC, Schneider_Collins}, has proven ideal for this purpose.  While a detailed description of the method must be relegated to the references, a few qualitative remarks here are useful.  The essential features of the FEDVR method are 1) to subdivide space into a variable number of elements of different length and 2) to use a flexible number of DVR basis functions in each element to accommodate the local  spatial variations of the wavefunction.
% based on the physics of that element.  
The basis functions in each element are chosen to be the Lagrange interpolation polynomials at the Gauss-Legendre-Lobatto quadrature points in that element. The Lobatto quadrature rule, which forces a common quadrature point at each boundary, enables the construction of a "bridge" basis function which ensures wavefunction continuity (communication) across the elements.  The "bridge" function is simply a linear combination of the last DVR function in region $i$ and the first DVR function in region $i+1$.  Note, that the order of the DVR basis in each element need not be the same. In addition, the use of a DVR basis in each element, means that matrix elements of the potential are simply the value of the potential at the Gauss-Legendre-Lobatto quadrature points in that element.  The components of the kinetic energy matrix in each element are given by analytic expressions involving the derivatives of the Lagrange interpolating polynomials evaluated at the quadrature points.  
%may be reduced to trivial analytic or numerical computation.  
First and higher order derivatives of the basis set are not and need not be continuous across the elements for the numerical procedure to be convergent in the limit of a complete basis.  The procedure enables us to capture both the short and long range features of the molecular wavefunctions with a reasonably small basis ($\sim$300 functions/channel). We thus have a large number of points in the potential region, where the wavefunction oscillates rapidly, and an increasingly sparse grid, as one gets deeper into the asymptotic region where the behavior of the wavefunction is governed purely by wavelength (energy) considerations.  Finally, an appealing attribute of the FEDVR basis for R-matrix calculations is that only a single FEDVR function contributes to the R-matrix elements at the boundary.  This is a consequence of the fact that the DVR functions are either zero or one at the FEDVR points.  

\section{Results and discussion}
\label{Results}

In this section we present results of calculations within our two-channel model, which fully illustrate the resonance scattering and molecular physics of the Feshbach resonance. In Table~\ref{parameters} the parameters defining the minimal two-channel model description and the set of physical parameters on which it is based are given for several Feshbach resonances of experimental interest. 
We present a comprehensive comparison of our two-channel results with those of a full coupled channels calculation only for the specific example of the Feshbach resonance occuring at $B_0$=202.15 G in fermionic $^{40}$K in the $|F=9/2,m_F=-7/2\rangle$ and $|F=9/2,m_F=-9/2\rangle$ hyperfine states. However, we have checked that our model is in agreement with more elaborate calculations for all the cases quoted in the table.
\begin{widetext}
%\begin{tabular}{|r|r|r|r|r|r|r|r|r|r|r|} \hline\hline
\begin{table}[htbp]
\begin{tabular}{|c|c|c|c|c|c!{\vrule width 1.5pt}c|c|c|c|c|} \hline\hline 
{} &
\multicolumn{5}{c!{\vrule width 1.5pt}}{Physical parameters} &
\multicolumn{5}{c|}{Model parameters} \\ \hline
Species & $C_6$ (au)  & $a_{\rm{bg}}$ (au) & $N_s$ & $B_0$ (G) & $\Delta B$ (G) & $\Delta\mu$ (MHz/G)
& $\epsilon$ (au) & $\sigma$ (au) & $W_0$ (au) & $E_{\rm{th}}$  (au) \\ \hline
$^{40}$K & 3897 & 172.06 & 27 & 202.15 & 7.788 & 2.35 & 1.631$\cdot 10^{-3}$ & 9.177 & 0.221 & 1.192$\cdot  10^{-7}$ \\ \hline
$^{40}$K & 3897 & 172.15 & 27 & 224.21 & 8.12 & 2.35 & 1.631$\cdot 10^{-3}$ & 9.177 & 0.226 & 1.112$\cdot  10^{-7}$ \\ \hline
$^{85}$Rb & 4700 & -480 & 35 & 155.199 & 10.641 & -3.345 & 1.124$\cdot 10^{-3}$ & 10.075 & 0.203 & 3.283$\cdot  10^{-7}$ \\ \hline
\hline
\end{tabular}
\caption{Physical parameters for a Feshbach resonance and the corresponding parameters describing the two-channel model. The units of the $C_6$ coefficient is Energy$\cdot$Length$^6$, with 1 au = 9.57344$\cdot 10^{-26}$ J nm$^6$. A frequency of 1 MHz corresponds to a temperature of 48 $\mu$K.}
\label{parameters}
\end{table}
\end{widetext}

\begin{figure}[htbp]
\begin{center}
\includegraphics[scale=0.4]{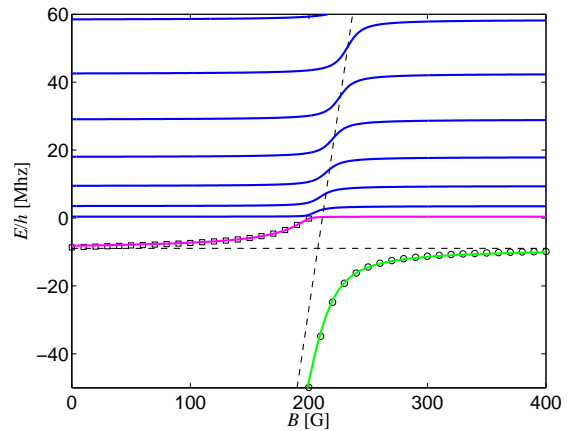}
\caption{Bound states and the discrete R-matrix representation of the scattering continuum for the $^{40}$K Feshbach resonance at $B_0=202.15$ G. The $E_{-1}$ (magenta line) and $E_{-2}$ (green line) bound states of the two-channel model as counted at $B=0$, are compared with the results of a full multi-channel calculation (open symbols). At the resonance position $B=B_0$ the number of bound states changes by one. The low energy R-matrix eigenstates with $E>0$ (blue lines) correspond to the eigenstates of a spherical box with vanishing derivative at the boundary. At higher energies no such correspondence exists (see text). The bare resonance state and the last bound state of the open channel potential with energy $E^{(1)}_{-1}=8.90$ MHz are indicated by the dashed lines.}
\label{E_multiple}
\end{center}
\end{figure}

The essential molecular physics of the Feshbach resonance is illustrated in Fig.~\ref{E_multiple}, 
where the energies of the last two bound states of the coupled channels problem are plotted (solid lines). The energies of the uncoupled diabatic states corresponding to the bare Feshbach molecule with slope $\Delta\mu$ and the last bound state of the open channel at energy $E^{(1)}_{-1}$ are indicated by the dashed lines.
In this case the coupling shifts the Feshbach resonance position downwards from the zero crossing of the bare Feshbash state at $B=B_{\rm{res}}$=211.39 G to $B_0$, where the binding energy of the last bound state at $B$=0 vanishes, and this state enters the continuum. The last bound state of the open channel is seen to play a pivotal role, as it interacts strongly with the bare resonance state, leading to an avoided crossing. Hence the dressed molecules produced in an adiabatic sweep of the magnetic field across the Feshbach resonance, starting at $B>B_0$, will evolve into the last bound state of the open channel potential (magenta curve) for $B\ll B_0$, not the bare resonance state in the closed channel. Instead it is the dressed state with energy $E_{-2}$ (green curve) which tracks $E_{\rm{res}}(B)$ in this regime. At fields much larger than $B_0$ the energy of this state approaches $E^{(1)}_{-1}$. This agrees with earlier observations by other authors~\cite{Szymanska2005}. For other resonances, such as the two in $^6$Li, the last bound state of the open channel lies much deeper, and has no effect on the Feshbach molecules.  

The open circles  in Fig.~\ref{E_multiple} give the results of a full coupled channels calculation. We observe that our simple two-channel model gives a faithful representation of the dressed states. The figure also shows the discrete R-matrix representation of the scattering continuum. At low energies they bear some semblance to the eigenstates in a spherical box with zero derivative on the boundary, though no such correspondence can be established at higher energies. Their spectral density is therefore tied to the size of the box. In particular, the first continuum state is not at exactly zero energy, but is displaced a small positive amount, representing the zero point energy in the effective box potential. The scattering resonance is visible as a scar across the continuum part of the spectrum. 

We remark that while the energy of the second to last adiabatic bound state approaches that of the bare resonance state far below the resonance, it is in general affected by the coupling to deeper bound states of the open channel potential and in fact never becomes linear in the magnetic field. It is therefore important to obtain the magnetic moment difference by fitting the position of the scattering resonance for fields above $B_0$. 

\begin{figure}[htbp]
\begin{center}
\includegraphics[scale=0.4]{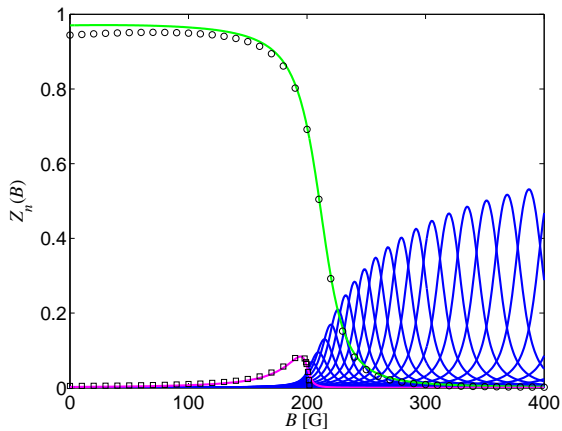}
\caption{Closed channel fraction for the energy curves in Fig.~\ref{E_multiple}. Note that for the molecular state created by a downward sweep across the Feshbach resonance the population in the closed channel never exceeds $8\%$. The closed channel population for the last two bound states resulting from a CC calculation are shown for comparison. Continuum curves with peak positions at higher magnetic fields correspond to R-matrix eigenstates with a higher energy.}
\label{Z_multiple}
\end{center}
\end{figure}

Figure~\ref{Z_multiple} gives additional details of the behavior of the dressed states as the magnetic field is varied. The figure displays the weight of the dressed state wavefunctions in the closed channel, defined as 
\begin{equation}
Z_n(B) = \int dr |\psi_{2,n}(r)|^2.
\end{equation}  
In accordance with the bound state spectrum in Fig.~\ref{E_multiple} the last adiabatic bound state is a pure open channel state in both the high and low field limit. Near the Feshbach resonance it acquires some closed channel character, but it remains dominated by the open channel. In contrast, the bound state with energy $E_{-2}$ changes from the open channel state with energy $E^{(1)}_{-1}$ to the bare resonance state in the closed channel, as the magnetic field is changed from above to below $B_0$~\cite{movies}. 
Again, the results of a coupled channels calculation are shown for comparison, and are found to confirm the simple model. For the coupled channels calculation $Z_n$ represents the sum of contributions from all the closed channels coupled with the open channel.
The variation of $Z_n$ with $B$ for the lowest lying states in the continuum shows that as the resonance energy increases with increasing magnetic field, the discrete Feshbach state becomes less diluted by the coupling to the continuum.  

\begin{figure}[htbp]
\begin{center}
\includegraphics[scale=0.4]{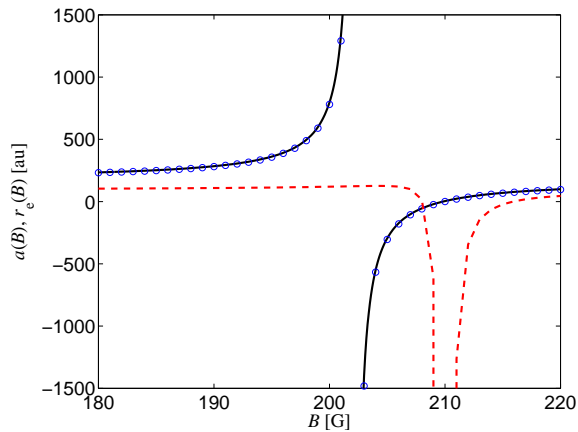}
\caption{Low energy scattering parameters for the two-channel model for the same resonance as above. The scattering length (blue open circles) and effective range (dashed line) found by fitting the energy dependent phase shift to the effective range expansion. The scattering length is compared with the resonance expression (\ref{a_vs_B}), shown by the solid line.}
\label{scattering}
\end{center}
\end{figure}

To extract information about the low energy scattering in the two-channel model, we find the scattering phase shift  as a function of energy using (\ref{tan_delta}). Fitting to the effective range expansion
$k\cot\delta(E)=-1/a(B)+r_e(B) k^2/2$ then gives the magnetic field dependent scattering length and effective range. These are plotted in Fig.~\ref{scattering}, and the scattering length is seen to be in agreement with (\ref{a_vs_B}), as expected since the two-channel model has been constructed to reproduce the salient features of this expression. For reference the effective range is also plotted. For most of the magnetic field interval shown $r_e$ remains small and positive, but it diverges at $B=B_0+\Delta B$=209.93 G, where the scattering length vanishes~\cite{Flambaum1999,Gao1998}. 

\begin{figure}[htbp]
\begin{center}
\includegraphics[scale=0.4]{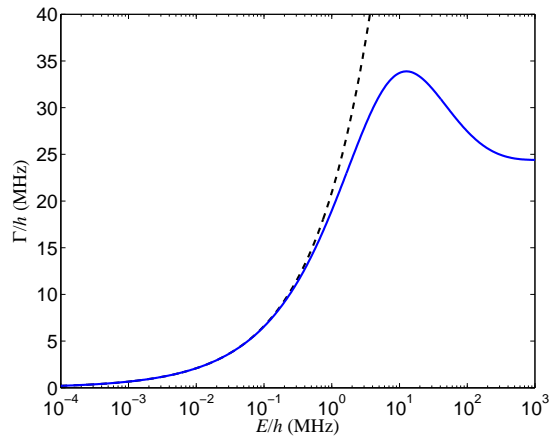}
\caption{Resonance width for $^{40}$K as a function of the scattering energy on a,logarithmic scale. The two-channel expression (\ref{Gamma}) (solid line) is compared to its limit in the Wigner threshold regime (\ref{Gamma_lowE}) (dashed line).}
\label{Ewidth}
\end{center}
\end{figure}

The energy width of the quasi-bound dressed Feshbach state as determined by (\ref{Gamma}) is plotted in Fig.~\ref{Ewidth}. As the energy approaches zero $\Gamma(E)\propto k$ as required by the Wigner threshold law. Using (\ref{DeltaB}) it follows that
\begin{equation}
\label{Gamma_lowE}
\lim_{k\rightarrow 0} \Gamma(E) = 2 a_{\rm{bg}} \Delta\mu \Delta B k.
\end{equation}
This Wigner threshold regime limit is also indicated in Fig.~\ref{Ewidth}.

\begin{widetext}
\begin{figure}
$\begin{array}{cc}
\includegraphics[scale=0.4]{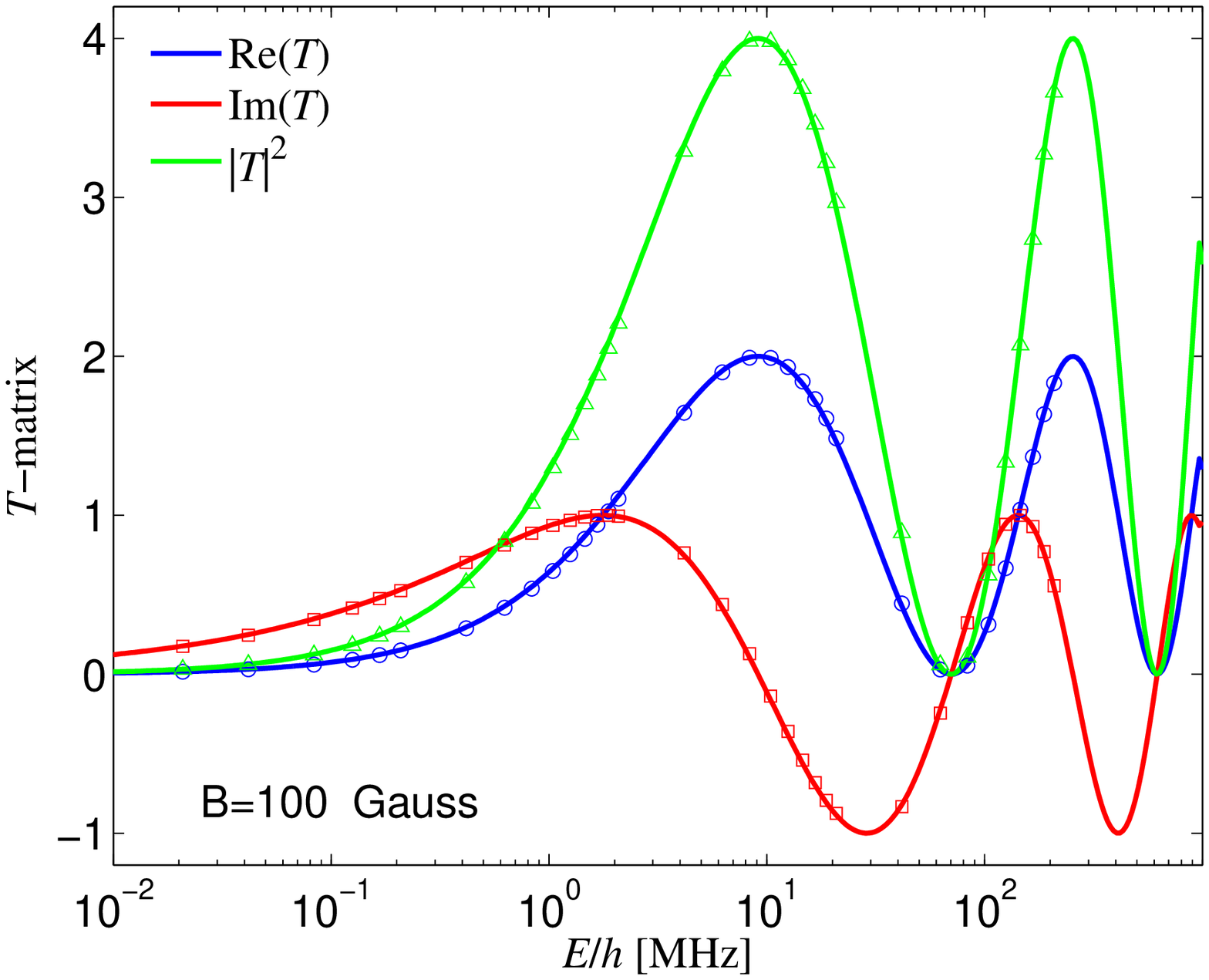} &
\includegraphics[scale=0.4]{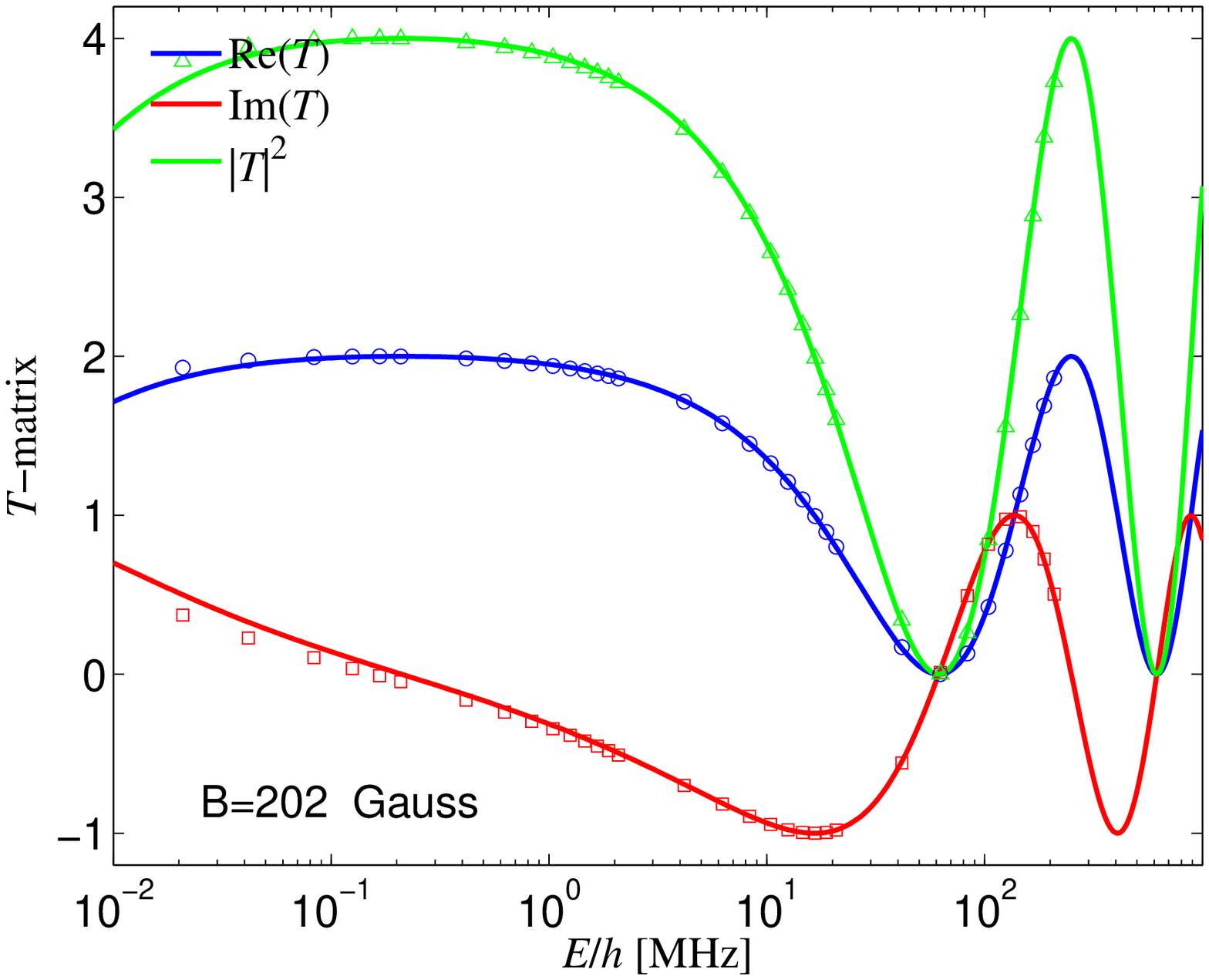}  \\
\includegraphics[scale=0.4]{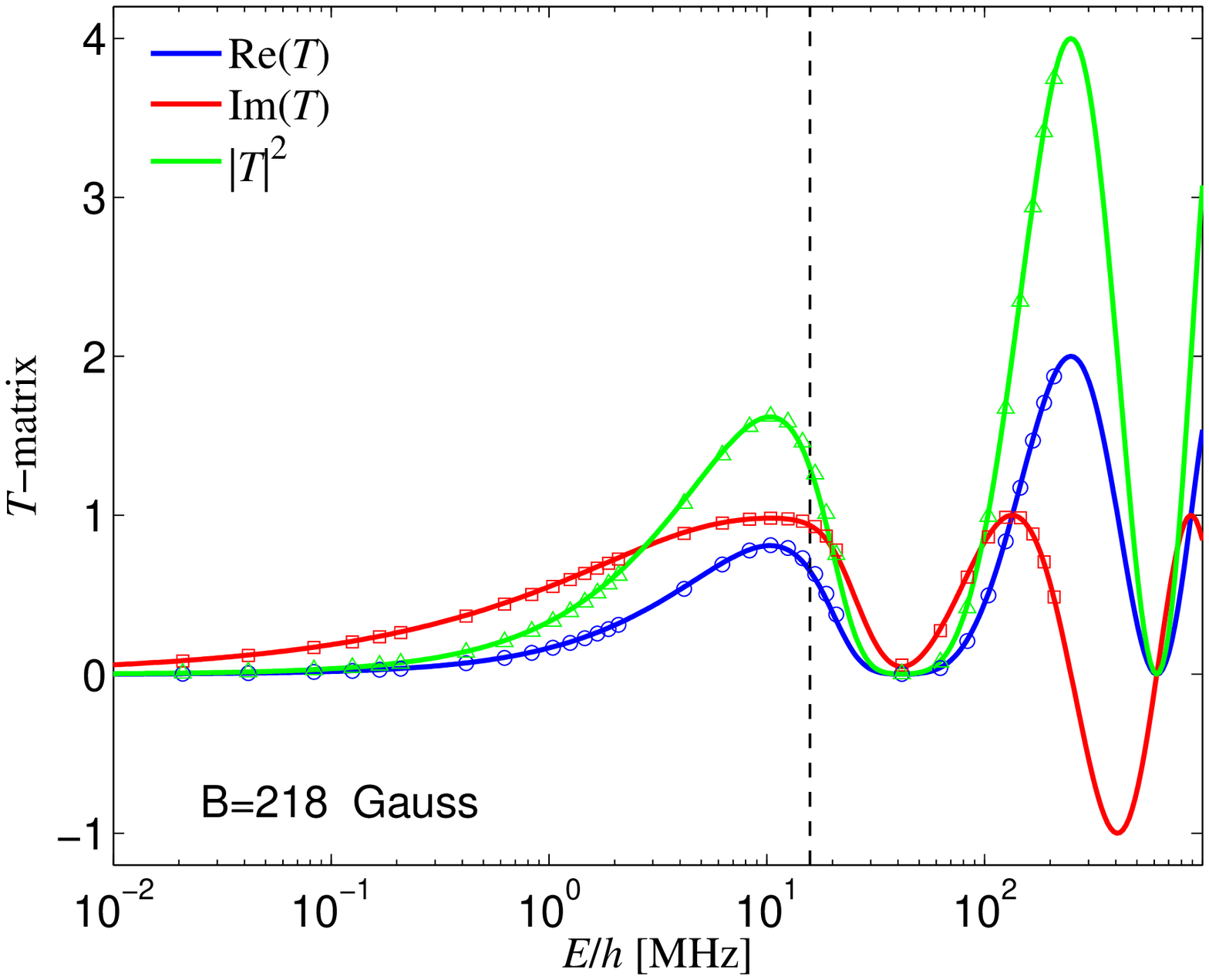} &
\includegraphics[scale=0.4]{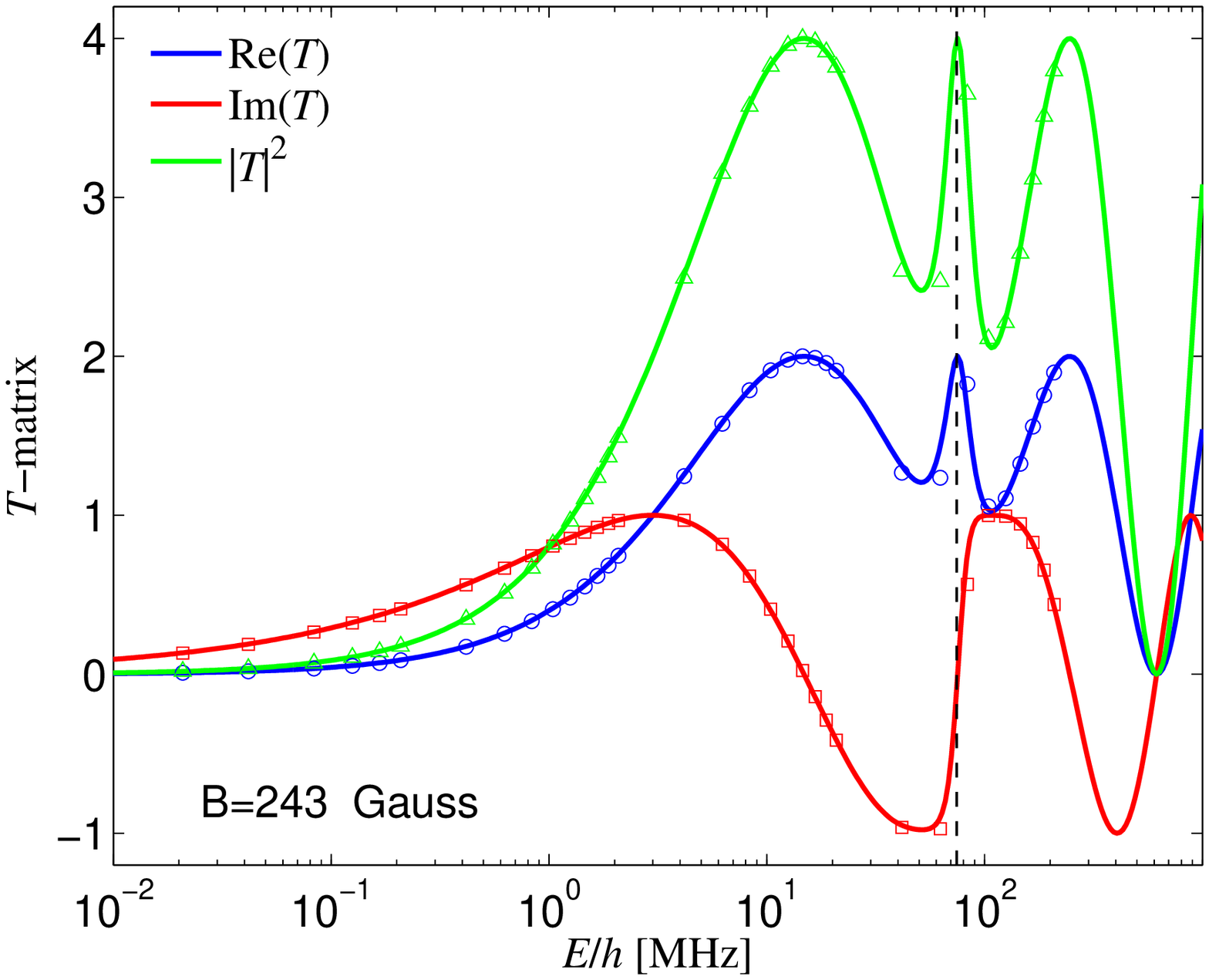} \\
\includegraphics[scale=0.4]{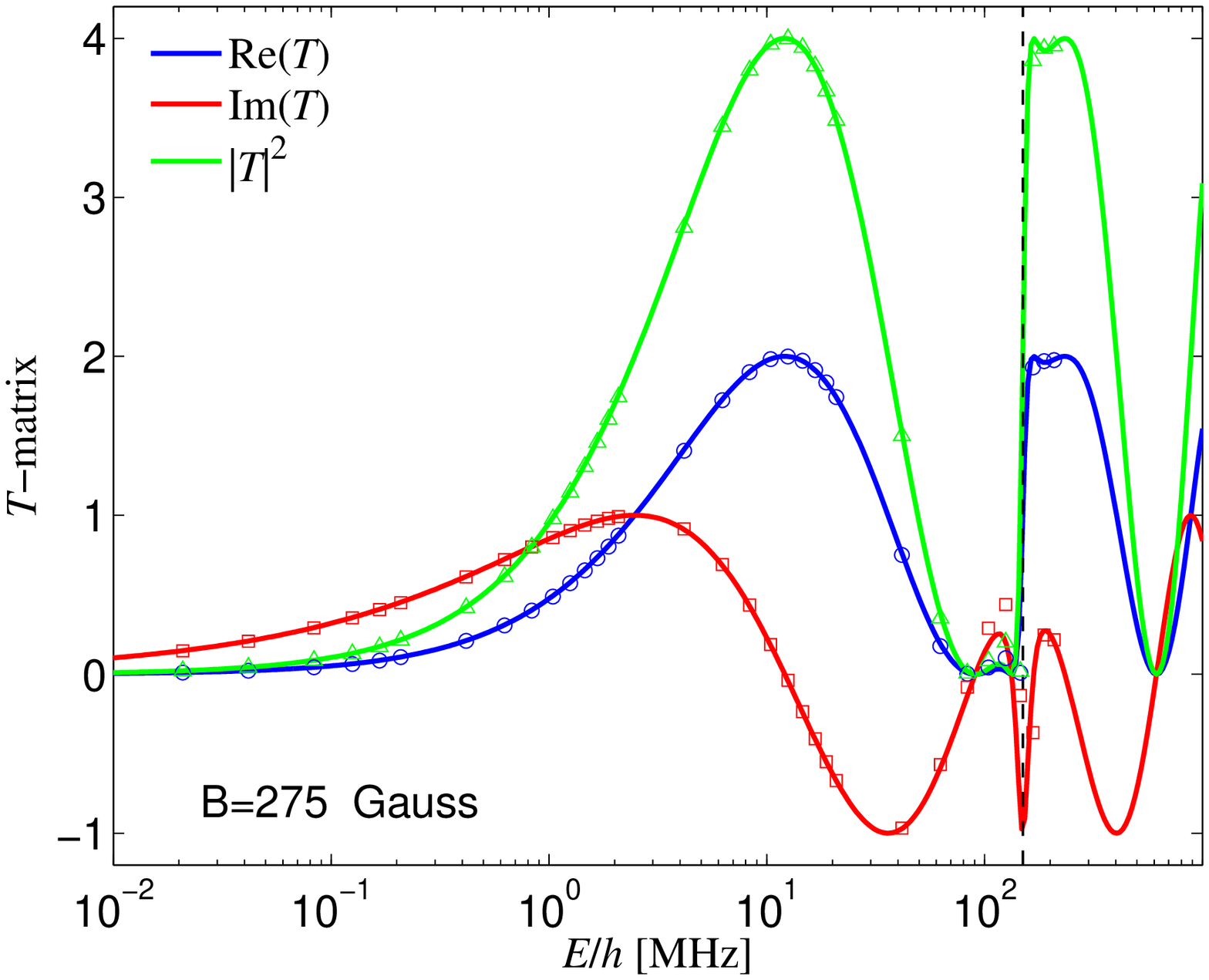} &
\includegraphics[scale=0.4]{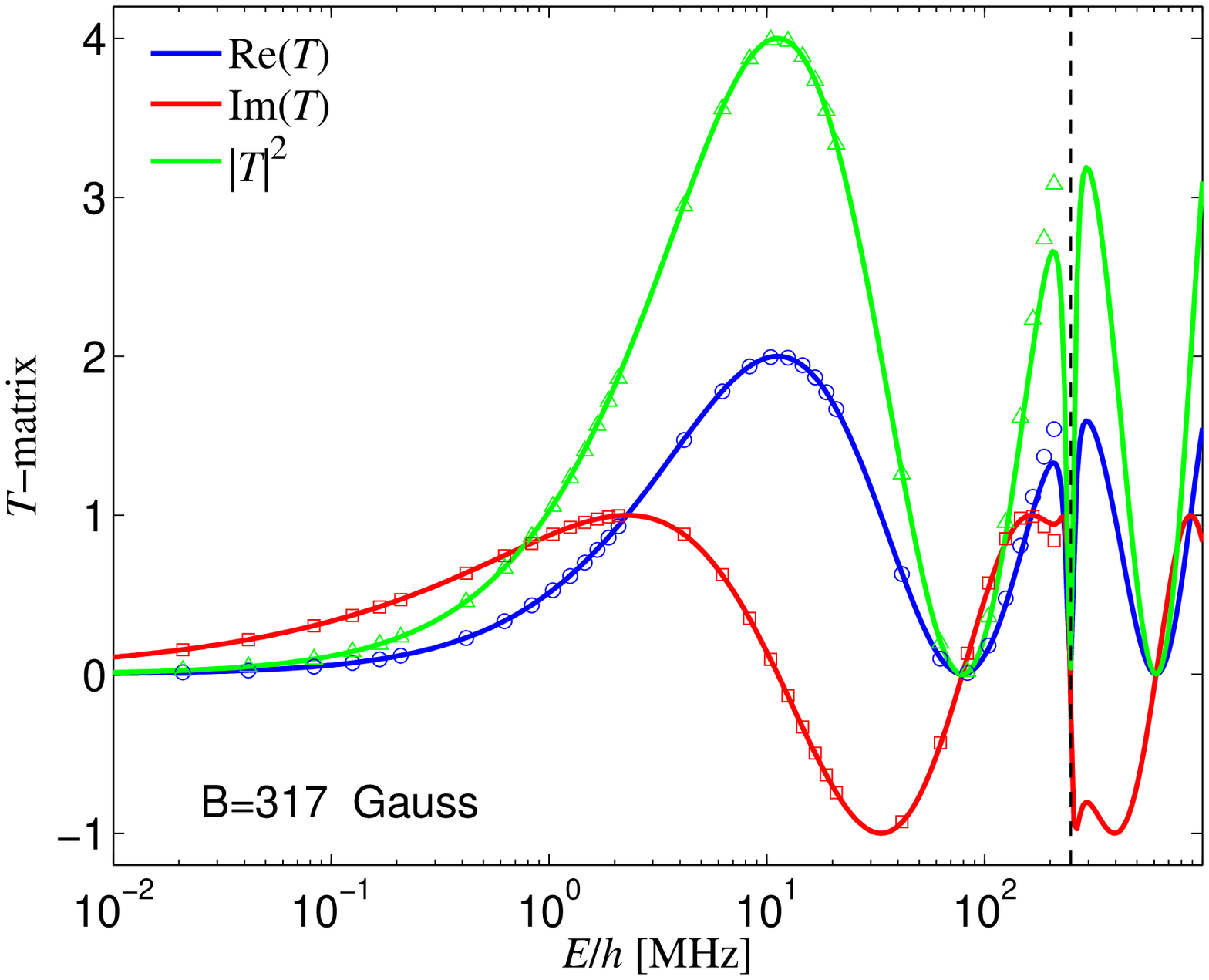} \\
\end{array} $
\caption{The energy variation on a logarithmic scale of the open channel component of the $T$-matrix for specified magnetic fields. The real and imaginary parts of $T_{11}(E)$ are shown in blue and red, respectively, while $|T_{11}|^2$ is plotted in green. For each magnetic field the results of  a full coupled channels calculation (open symbols) are compared with those of the two-channel model (full lines). For fields higher than $B_{\rm{res}}$ the energy of the bare resonance state is indicated by the vertical dashed line. The first panel corresponds to background scattering, while in the second panel the magnetic field is tuned to the Feshbach resonance. For $B=243$ G, $B=275$ G, and $B=317$ G the background phase shift (modulo $\pi$) is close to $0$, $\pi/4$, and $\pi/2$, respectively (compare with the plot for $B=100$ G)~\cite{movies}.}
\label{T_matrix}
\end{figure}
\end{widetext}

To illustrate in greater detail how the scattering evolves with energy and magnetic field we plot in Fig.~\ref{T_matrix} the energy dependent $T$-matrix, ${\bf T}={\bf I}-{\bf S}$, for several different values of the magnetic field. We only plot the purely open channel component of ${\bf T}$, which is related to the phase shift via 
\begin{equation}
|T_{11} (E,B)|^2 = 4 \sin^2(\delta(E,B)).
\end{equation}
The total phase shift may be written as a sum of a background and the resonant contribution~\cite{Feshbach1958_1962,Fano1961}
\begin{equation}
\delta(E)=\delta_{\rm{bg}}(E)+\delta_{\rm{res}}(E,B). 
\end{equation}
For the resonance under consideration here we may take the background term to be independent of the magnetic field, but this is not always the case (see {\textit{e.g.}}~\cite{Bartenstein2005}).
For each magnetic field the figure compares the energy variation of both the real and imaginary parts of $T_{11}$ calculated within the two-channel model of this paper and full coupled channels calculations. For completeness we also plot for both models the resulting $|T_{11}|^2$. At low values of $B$ the Feshbach resonance has almost no influence on the scattering, which is dominated by the background contribution. Right on the Feshbach resonance the scattering near threshold is strongly affected by the coupling to the closed channel bound state. In the last four panels the energy of the bare resonance state is indicated with a vertical dashed black line. When it lies in the continuum it gives rise to a scattering resonance, the shape of which depends value of the background phase shift in the neighborhood of the resonance position. At $B=218$ G the resonance is broad (corresponding to a finite value of ${\rm{Im}}(T_{11})$), and its position does not line up with the energy of the bare resonance state. At higher fields the resonance follows the bare resonance state. 

The last three panels demonstrate the interplay between the two different contributions to the scattering by plotting $T_{11}$ at magnetic field values, where the resonance occurs at energies such that the background phase shift modulo $\pi$ is close to $0$, $\pi/4$, and $\pi/2$, respectively. As the scattering energy is increased through the resonance the phase shift increases by approximately $\pi$. Correspondingly, the values of the real and imaginary parts of $T_{11}$ reflect the value of the background phase shift on resonance. This gives ${\rm{Re}}T_{11}(B=243 \ {\rm{G}}) \approx 2$, ${\rm{Re}}T_{11}(B=275 \ {\rm{G}}) \approx 1$, and ${\rm{Re}}T_{11}(B=317 \ {\rm{G}}) \approx 0$ for the real part of the $T$-matrix on resonance, while the imaginary part at the same magnetic fields is 
${\rm{Im}}T_{11}(B=243 \ {\rm{G}}) \approx 0$, ${\rm{Im}}T_{11}(B=275 \ {\rm{G}}) \approx -1$, and ${\rm{Im}}T_{11}(B=317 \ {\rm{G}}) \approx 0$,
in good agreement with the numerical findings.

\section{Conclusions}
\label{Conclusions}

We have presented a recipe for constructing a simplified two-channel model of an isolated Feshbach resonance. Given a set of five physical parameters, which can be obtained from either experiments or exact calculations, the parameters characterizing our model can be derived without ambiguity.  Within the R-matrix formalism, the bound state part of the spectrum as well as the states scattering at arbitrary energies in the continuum are obtained from a diagonalization of the two-channel Hamiltonian (including the Bloch operator). The FEDVR is particularly well suited to represent the model Hamiltonian, since the combination of short range molecular potentials with long wave length scattering requires a spatial grid which encompasses two widely separated length scales.  

We have provided the parameters defining the two-channel model for several Feshbach resonances  used in current experiments, and made direct comparison between the predictions of our model with the results of exact calculations. This comparison shows that over a large range of energies and magnetic fields the two-channel model reliably reproduces the results of full coupled channels calculations. The usefulness of the two-channel model lies in its simplicity. In particular,  the magnetic field enters only as a parameter in the closed channel potential. Hence the stationary two-channel Hamiltonian provides an excellent starting point for studies of propagation in the presence of a time dependent magnetic field. Such time-dependent problems are also well suited for the FEDVR method~\cite{Schneider_Collins}. 

We end by contrasting the two-channel scheme of this paper with simplified models of Feshbach resonances developed by other authors. There are examples of a number of similar two-channel models in the literature~\cite{Mies2000,Koehler2003,Goral2004,Koehler2004,Julienne2004}. The new elements of this work are the use of pseudopotentials, which gives as realistic a representation of the molecular potentials of the full coupled channels problem as possible, and a detailed prescription for generating them. Specifically, we choose model potentials which support the same number of bound states as the entrance channel in the full problem. We do not anticipate that the additional details of the model presented here will lead to fundamentally different results. But 
it does lead to a more accurate representation of the dressed state wavefunctions. Similar remarks apply when our model is compared with those in which even simpler pseudopotentials are introduced, such as a well with a barrier~\cite{DePalo2004}, two coupled wells~\cite{Kokkelmans2002,Chin2005} or zero range potentials~\cite{Moore2005}. These approaches have the advantage that they are amenable to analytical treatment, but they give no details of the molecular physics at short range.

The virtue of a model description is measured on its ability to reproduce the available data effectively, while illuminating the important underlying physical principles. We believe that our approach coupled with an R-matrix analysis implemented within a FEDVR scheme constitutes an efficient and transparent procedure for generating the bound state and scattering properties of the coupled channels Feshbach problem.

\clearpage

\clearpage


\begin{thebibliography}{99}

\bibitem{Feshbach1958_1962} H. Feshbach, Ann. Phys. (N.Y.) {\bf{5}}, 357 (1958); {\bf{19}}, 287 (1962).
\bibitem{Tiesinga1993} E.~Tiesinga, B.~J.~Verhaar and H.~T.~C.~Stoof, Phys. Rev. A {\bf{47}}, 4114 (1993).
\bibitem{Moerdijk1995} A.~J.~Moerdijk, B.~J.~Verhaar and A.~Axelsson, Phys. Rev. A {\bf{51}}, 4852 (1995).
\bibitem{Stoof1988} H.~T.~C.~Stoof, J.~M.~V.~A.~Koelman and B.~J.~Verhaar, Phys. Rev B {\bf{38}}, 4688 (1988).
\bibitem{Ohara2002}  K.~M.~O'Hara, S.~L.~Hemmer, M.~E.~Gehm, S.~R.~Granade and  J.~E.~Thomas, Science {\bf{298}}, 2179 (2002).
\bibitem{Regal2004} C.~A.~Regal, M.~Greiner and D.~S.~Jin, Phys. Rev. Lett. {\bf{92}}, 040403 (2004).
\bibitem{Zwierlein2004} M. W. Zwierlein, C. A. Stan, C. H. Schunck, S. M. F. Raupach, A. J. Kerman and W. Ketterle, Phys. Rev. Lett. {\bf{92}},  120403 (2004).
\bibitem{Bartenstein2004} M.~Bartenstein, A.~Altmeyer, S.~Riedl, S.~Jochim, C.~Chin, J.~Hecker Denschlag and R.~Grimm, Phys. Rev. Lett. {\bf{92}}, 120401 (2004).
\bibitem{Kinast2004} J.~Kinast, S.~L.~Hemmer, M.~E.~Gehm, A.~Turlapov, and J.~E.~Thomas, Phys. Rev. Lett. {\bf{92}}, 150402 (2004).
\bibitem{Bourdel2004}  T.~Bourdel, L.~Khaykovich, J.~Cubizolles, J.~Zhang, F.~Chevy, M.~Teichmann, L.~Tarruell, S.~J.~J.~M.~F.~Kokkelmans and C.~Salomon, Phys. Rev. Lett. {\bf{93}}, 050401 (2004). 
\bibitem{Goral2004} K. G{\'{o}}ral, T. K{\"{o}}hler, S. Gardiner, E. Tiesinga and P. Julienne, J. Phys. B {\bf{37}}, 3457 (2004).
\bibitem{Koehler2004} T.~K{\"o}hler, K.~G{\'{o}}ral and T.~Gasenzer, Phys. Rev. A {\bf{70}}, 023613 (2004).  
\bibitem{Simonucci2005} S.~Simonucci, P.~Pieri and G.~C.~Strinati, EuroPhys. Lett. {\bf{69}}, 713 (2005).
\bibitem{Szymanska2005} M.~H.~Szyma{\'{n}}ska, K.~G{\'{o}}ral, T.~K{\"{o}}hler and K.~Burnett, Phys. Rev. A {\bf{72}}, 013610 (2005).
\bibitem{Stoll2005} M.~Stoll and T.~K{\"o}hler, Phys. Rev.  A {\bf{72}}, 022714 (2005).  
\bibitem{Mies2000} F.~H.~Mies, E.~Tiesinga and P.~S.~Julienne, Phys. Rev. A {\bf{61}}, 022721 (2000). 
\bibitem{Koehler2003} T.~K{\"o}hler, T.~Gasenzer, P.~S.~Julienne and K.~Burnett, Phys. Rev. Lett. {\bf{91}}, 230401 (2003).
\bibitem{Fano1961} U. Fano, Phys. Rev. {\bf{124}}, 1866 (1961).
\bibitem{Flambaum1999} V.~V.~Flambaum, G.~F.~Gribakin and C.~Harabati, Phys. Rev. A {\bf{59}}, 1998 (1999).
\bibitem{P.Burke_C.Noble_V.Burke} A comprehensive and up to date article, to be published in Advances in Atomic and Molecular Physics is, P. G. Burke, C. J. Noble and V. M. Burke, "R-Matrix Theory of Atomic, Molecular and Optical Processses"
\bibitem{B.Schneider_R-Matrix} B. I. Schneider, "The R-Matrix Theory of Electron-Molecule Scattering", in, Electron-Atom and Electron-Molecule Collisions, Ed. by J. Hinze, Plenum Press, New York and London (1983)
\bibitem{Julienne2004} P.~S.~Julienne, E.~Tiesinga and T.~K{\"o}hler, J. Mod. Optics {\bf{51}}, 1787 (2004). 
\bibitem{Bloch} C. Bloch, Nucl. Phys. {\bf{4}}, 503 (1957).
\bibitem{Mott1965} N.~F.~Mott and H.~S.~W.~Massey, {\textit{The Theory of Atomic Collisions}}, 3rd ed. (Oxford University Press, Oxford, 1965).
\bibitem{WigEis} E. Wigner and L. Eisenbud, Phys. Rev. {\bf{72}}, 29 (1947)
\bibitem{Buttle} P. J. A. Buttle, Phys. Rev. {\bf{160}}, 719 (1967)
\bibitem{Siegert} O.~I.~Tolstikhin, V.~N.~Ostrovsky and H.~Nakamura,  Phys. Rev. A {\bf{58}}, 2077 (1998).
\bibitem{ResMcC} T. N. Rescigno and C. W. McCurdy, Phys. Rev. A {\bf 62}, 032706-1 (2000)
\bibitem{Schneider_Collins} B.~I.~Schneider and L.~Collins, J. Non-Cryst. Solids {\bf{351}}, 1551 (2005).
\bibitem{movies} See EPAPS Document No. [{\textit{number will be inserted by publisher}}] for animations showing the variation of the open and closed channel components of the $n=-2$ and $n=-1$ dressed state wavefunctions with magnetic field, as well as the magnetic field dependence of the $T$-matrix. This document can be reached via a direct link in the online article's HTML reference section or via the EPAPS homepage (http://www.aip.org/pubservs/epaps.html).
\bibitem{Gao1998} B.~Gao, Phys. Rev. A {\bf{58}}, 4222 (1998).
\bibitem{Bartenstein2005} M.~Bartenstein, A.~Altmeyer, S.~Riedl, R.~Geursen, S.~Jochim, C.~Chin, J.~Hecker Denschlag, R.~Grimm, A.~Simoni, E.~Tiesinga, C.~J.~Williams and P.~S.~Julienne, Phys. Rev. Lett. {\bf{94}}, 103201 (2005). 
\bibitem{DePalo2004} S.~De Palo, M.~L.~Chiofalo, M.~J.~Holland and S.~J.~J.~M.~F.~Kokkelmans, Phys. Lett. A {\bf{327}}, 490 (2004).
\bibitem{Kokkelmans2002} S.~J.~J.~M.~F.~Kokkelmans, J.~N.~Milstein, M.~L.~Chiofalo, R.~Walser and M.~J.~Holland, Phys. Rev. A {\bf{65}}, 053617 (2002).
\bibitem{Chin2005} C.~Chin, cond-mat/0506313 (2005).
\bibitem{Moore2005} M.~G.~Moore, cond-mat/0506383 (2005).
\end{thebibliography}
\end{document}